\documentclass[conference]{IEEEtran}
\IEEEoverridecommandlockouts
\usepackage{cite}
\ifCLASSINFOpdf
  \usepackage[pdftex]{graphicx}
  \DeclareGraphicsExtensions{.pdf,.jpeg,.jpg,.png}
\else
  \usepackage[dvips]{graphicx}
  \DeclareGraphicsExtensions{.eps,.jpeg,.jpg,.png}
\fi
\usepackage{amsmath}
\usepackage{amssymb}
\usepackage{array}

\usepackage{todonotes}
\usepackage{multirow}

\begin{document}

\title{Capacity Value of Interconnection\\Between Two Systems}

\author{\IEEEauthorblockN{Simon H. Tindemans}
\IEEEauthorblockA{Department of Electrical Sustainable Energy\\
Delft University of Technology\\
The Netherlands\\
s.h.tindemans@tudelft.nl}
\and
\IEEEauthorblockN{Matthew Woolf and Goran Strbac\thanks{This work has received funding from the UK Energy Research Centre (UKERC) as part of the Future Energy System Pathways programme.}}
\IEEEauthorblockA{Department of Electrical and Electronic Engineering\\
Imperial College London\\
United Kingdom\\
\{matthew.woolf, g.strbac\}@imperial.ac.uk}}


\maketitle

% SIMON: Request loading of control commands from the additional bibliography
\bstctlcite{IEEE:BSTcontrol}

\begin{abstract}
Concerns about system adequacy have led to the establishment of capacity mechanisms in a number of regulatory areas. Against this background, it is essential to accurately quantify the contribution to security of supply that results from interconnectors to neighbouring systems. This paper introduces a definition of capacity value for interconnection between two systems in the form of a capacity allocation curve. Four power flow policies are proposed to encompass the full range of possible market outcomes that may affect the capacity value. A convolution-based method is presented to efficiently compute and compare capacity allocation curves, and it is applied to a model system that is inspired by Great Britain's interconnection with the continental Europe. The results indicate areas of interest for the coordination of capacity mechanisms. 
\end{abstract}


% \fi
%\IEEEpeerreviewmaketitle

\section{Introduction}

Electricity networks are increasingly interconnected. As an example of this development, at the time of writing the Great Britain system has a total interconnection capacity of 4~GW (3~GW to continental Europe), with a further 7.7~GW in development until 2021. %\footnote{https://www.ofgem.gov.uk/electricity/transmission-networks/electricity-interconnectors}. 
The European Commission has proposed a target of 15\% import capacity over installed generation capacity in each European Union Member State \cite{EuropeanCommissionExpertGrouponelectricityinterconnectiontargets2017TowardsTargets}. 
Price arbitrage is the primary driver of this anticipated expansion, but it is well established that interconnection also enhances security of supply through the sharing of available generation assets between neighbouring systems. 

In recent years, concerns about security of supply in systems that increasingly rely on renewable generation sources and long-distance transmission have led to the introduction of capacity mechanisms. The extent to which an asset (generator, storage facility, interconnector) participates in these capacity mechanisms is determined by its capacity value, which expresses the effect of the asset on the security of supply in the `common currency' of an equivalent change in firm load or generation level \cite{Zachary2011}. There has been substantial interest in defining capacity values for different assets, including tidal barrages \cite{Radtke2011CapacityBarrages}, wind power \cite{Keane2011CapacityPower}, solar power \cite{Duignan2012CapacityPower}, energy storage \cite{Sioshansi2014AStorage} and distributed generation \cite{Dent2015DefiningGeneration}. 

Extending this analysis to transmission assets, in particular interconnection between two regulatory areas, requires addressing two challenges. The first is the uncertain realization of power flows between two systems, amid concerns that economic factors may cause interconnectors to export at times of insufficient system margins. Interconnector owners have been permitted to participate in the Great Britain (GB) capacity auctions since December 2015 (T-4 auction for 2019/20 delivery). The GB transmission system operator, National Grid, currently uses a variety of economic approaches (historical extrapolation and modelling) to determine the capacity value \cite{2018NationalReport} of interconnection.

The second challenge is that the contribution of interconnection to security of supply to a single system cannot be determined in isolation, because capacity decisions in neighbouring systems are related. Against this background, the European Commission is investigating the possibility of an EU-wide probabilistic adequacy assessment methodology that would enable a consistent assessment of security contributions across borders \cite{2016FinalMechanisms}. 

Quantifying the security of supply contribution of interconnection is a research topic with a long history; see e.g. 
\cite{Cook1963DeterminationSystems,Vassell1972AnalysisSystems,Pang1975}. However, to the authors' knowledge, no formal definition of capacity value has been given in this setting, and results have focused on specific problem settings and computational challenges rather than illustrating trade-offs involved in coordinated operation of capacity mechanisms. 

In this paper, we introduce a definition of capacity value for interconnection between two systems in the form of a capacity allocation curve. Instead of an explicit model of market driven power flows, a set of four policies is used to encompass the full range of possible market outcomes in order to determine the capacity value in a robust manner. A convolution-based method is presented to efficiently compute and compare capacity allocation curves, and it is applied to a model system that is inspired by Great Britain's interconnection with continental Europe. The results indicate areas of interest for the coordination of capacity mechanisms.  

\section{Capacity value definition}

We consider two interconnected systems, labeled A and B\footnote{We use `systems' and `interconnectors' instead of `zones' and `tie lines'. The terminology is interchangeable as long as each system/zone is associated with a local risk measure, and not merely a part of one larger system.}. 
In this paper, we shall assume that the loss of load risk in each of these systems is measured using the LOLE risk measure as it is commonly used in Europe: the expected number of hours per year in which a system is unable to supply all load. For a generation adequacy model with hourly resolution, the LOLE is effectively the loss of load probability (LOLP) across all hours and system states, expressed in hours per year. In the absence of interconnection, the LOLE is then defined as
\begin{equation}
r^0_x = h \times \textrm{LOLP} = h \times \textrm{Pr}(M_x < 0),
\end{equation}
where $h$ is the number of hours per year ($8760$ in a non-leap year), $x \in {A,B}$ is the system index and $M_x$ is a random variable representing the net generation margin (generation minus demand) in system $x$. To simplify notation in what follows, we shall use $r^0$ to represent the 2-vector $\left( r^0_A, r^0_B \right)^T$.  

In the context of interconnection between two independently operated systems, each system operator is responsible for security of supply in the local system. The reference risk $r^0$ represents the baseline case where each system disregards interconnection capacity for adequacy assessment purposes. However, in practice, both systems can benefit from the exchange of power over interconnecting lines. In order to define the capacity value concept of interconnection, we generalize the concept of Equivalent Load Carrying Capacity (ELCC)\cite{Zachary2011}. Qualitatively, the ELCC is the amount of additional (constant) load that the system can support by using the interconnection between systems A and B - without sacrificing security of supply with respect to the baseline.

We opt for an ELCC-based metric in favour of the similar EFC-based metric, because it is well-suited to the problem at hand. We aim to quantify the overall benefit of accurately incorporating interconnection in a capacity mechanism, compared to a reference case where each system builds sufficient capacity in isolation. Due to the mathematical identity between the addition of constant load and a reduction of firm (100\% available) generation capacity, we may consider the ELCC equivalent to the amount of firm capacity that can be retired (or not purchased in a capacity auction). 

In order to formally define a capacity value of interconnection, we define the interconnection-adjusted risk as follows:
\begin{equation}
r^+_x(l) = h \times \textrm{Pr}[M'_x(l) < 0]. \label{eq:rplus}
\end{equation}
Here, the random variable $M'_x$ is the generation margin in system $x$ when making use of interconnection, and $l = (l_A, l_B)^T$ is a 2-vector of constant demand added to both systems. The relation between the unadjusted margin $M_x$ and the adjusted margin $M'_x(l)$ depends on the interconnection properties and operational policies, and will be discussed in the next section. We shall assume that $r^+_x(l)$ is a monotonically increasing function in each coordinate of $l$ (i.e. adding load never increases security of supply). 

\begin{figure}[!t]
\centering
\includegraphics{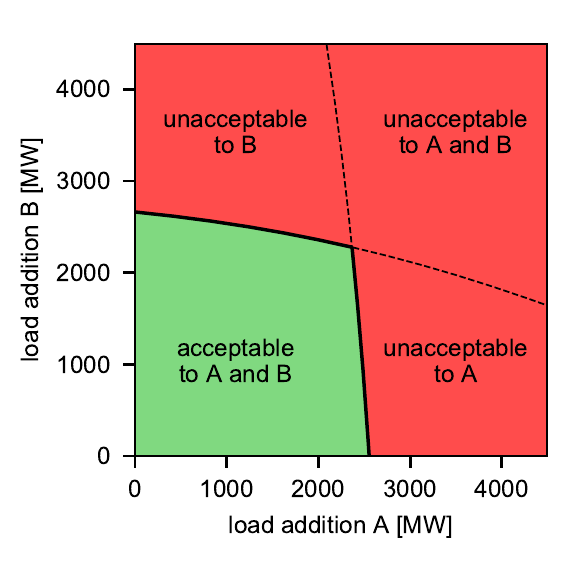}
\caption{Illustration of the acceptable set $\mathcal{A}$ (in green) and the capacity allocation curve (bold black), corresponding to the 3~GW interconnection scenario in Fig.~\ref{fig:allocation}. Load additions in the red areas cause risks to exceed baseline levels $r^0$ in system A, B or both.}
\label{fig:acceptableset}
\end{figure}

The set of \emph{acceptable load additions} -- additions that result in risks that do not exceed the baseline risk -- is defined as\footnote{We assume that $r^+_x(l)$ is a continuous function of $l$; the discontinuous case can be handled -- at the expense of notational complexity -- by defining $\mathcal{A}$ as the closure of the set of acceptable load additions.}
\begin{equation}
\mathcal{A} = \{ l \in \mathbb{R}^2 : r^+_A(l) \le r^0_A, r^+_B(l) \le r^0_B \}. \label{eq:setAcceptable}
\end{equation}
The ELCC is intuitively defined as the largest such load, but $\mathcal{A}$ does not have a unique largest element. Hence, we define the Pareto optimal \emph{ELCC set} as follows:
\begin{equation}
\mathcal{C}_p = \{  l \in \mathcal{A} : \nexists \ l'  \in \mathcal{A}: l'\neq l, l'_A \ge l_A,  l'_B \ge l_B  \}. \label{eq:ELCCset}
\end{equation}
In general, this set is not connected, which can be a disadvantage for numerical analysis and visualization. For this reason, we slightly relax the notion of optimality to obtain a connected set, by including load vectors that are equal in one of their components. This results in the \emph{capacity allocation curve}
\begin{equation}
\mathcal{C} = \{  l \in \mathcal{A} : \nexists \ l'  \in \mathcal{A}: l'_A > l_A,  l'_B > l_B  \}. \label{eq:capcurve}
\end{equation}
The construction of the acceptable set and the capacity allocation curve are illustrated in Fig.~\ref{fig:acceptableset}. In practical applications, a single optimal value may be selected from the capacity allocation curve set by defining a value (or cost) function $v(l) \in \mathbb{R}$. For example, the value function may be the sum of the load components (for a max-sum objective), or it could reflect the financial benefits resulting from capacity reductions in both systems. Under a value function that is monotonic in both parameters, there is no difference between using the ELCC set \eqref{eq:ELCCset} and the capacity allocation curve \eqref{eq:capcurve}.

\section{Determining combined system margins}

The capacity allocation curve is calculated from the adjusted capacity margin $M' = (M'_A, M'_B)^T$ using \eqref{eq:rplus}. This section shows how $M'$ is determined from the baseline margin $M$. This requires defining an operational policy for the power flows on the interconnecting lines and an appropriate computational procedure.

\subsection{Power flow policies}

The real-time flow of power between A and B is subject to market forces that will not be modelled. We shall only make the safe assumption that whenever an adequate solutions are available (i.e. non-negative margins in A and B), market and regulatory incentives are such that all load is supplied. This leaves undefined only those cases where a shortfall cannot be resolved to each area's satisfaction. Four policies are implemented that collectively span the range of possible market outcomes:
\begin{itemize}
\item \textbf{Veto:} Each area only exports power if it has a net generation surplus. This policy guarantees that the security contribution of electrical interconnection is non-negative. This is a common assumption in multi-area studies \cite{Billinton1996}, but it should be noted that such export restrictions are incompatible with the free operation of a multi-area market, and with the EU's Internal Energy Market in particular \cite{2016FinalMechanisms}.
\item \textbf{Share:} Areas share any shortfalls in proportion to their demand levels, insofar as this is possible given the available interconnection capacity. This policy can be thought of as the outcome of real-time market clearing where prices increase in proportion to the fraction of load shed in each system. 
\item \textbf{Assist A:} Interconnection capacity is used to eliminate shortfalls within system A to the maximum extent possible, even if this causes shortfalls in system B.
\item \textbf{Assist B:} The reverse of \emph{Assist A}. 
\end{itemize}

\subsection{Margin computation}
The calculation of capacity value curves will be performed using a convolution method, which allows for an explicit enumeration of system margins and their availabilities without the uncertainties that are inherent in a Monte Carlo approach. Similar enumeration methods were employed in a multi-area setting in \cite{Vassell1972AnalysisSystems, Pang1975}.
Despite the inherent limitations of enumeration approaches in dealing with complex systems, they are well suited to an investigation of generic properties of the capacity value of interconnection. 

We consider model systems where the generation margin is determined by three components: (i) available capacity from conventional generation, (ii) wind power and (iii) demand. These are represented by two-dimensional random variables $G$, $W$ and $D$, respectively, defining the two-dimensional system margin as
\begin{equation}
M = G + W - D.
\end{equation}
The random variables $G$, $W$ and $D$ are assumed to be mutually independent. In addition, the available capacity from thermal generation in system A ($G_A$) is assumed to be independent from that in system B ($G_B$), allowing a separable representation. The probability density function $f_M$ of $M$ can therefore be computed by convolution as
\begin{equation}
f_M(m_A, m_B) = (f_{G_A} \star f_{G_B} \star f_{W} \star f_{-D})(m_A, m_B). \label{eq:convolve}
\end{equation}

\subsection{Margin adjustment}

The baseline margin $M$ must be adjusted to reflect the addition of load $l$ and the interconnection between both systems, subject to the operational policy. As an intermediate step to calculating the risk $r_x^+(l)$ as defined in \eqref{eq:rplus}, we calculate the loss of load probability (LOLP) in system A for a given interconnection capacity $c$:
\begin{equation}
p_A(l;c) = \textrm{Pr}[M'_A(l) < 0|c]. \label{eq:defrho}
\end{equation}
We define the probability
\begin{equation}
\phi_A(l_A, l_B) = \int_{-\infty}^{l_A} \int_{-\infty}^{l_B + (l_A - m_A)} f_M(m_A, m_B) \mathrm{d}m_B \mathrm{d}m_A, \label{eq:integralphi}
\end{equation}
which integrates the margin $f_M$ over an area defined by the inequalities $m_A \le l_A$ and $m_A + m_B \le l_A + l_B$. We identify three non-overlapping integrals of $f_M(\cdot)$ that stand for three distinct contributions to the shortfall probability:
\begin{align}
p_{A}^{\textrm{base}}(l;c) = & \phi_A(l_A -c, \infty) \label{eq:base} \\
p_{A}^{\textrm{imp}}(l;c) =& \phi_A(l_A, l_B) - \phi_A(l_A-c, l_B+c) \label{eq:import} \\
p_{A}^{\textrm{exp}}(l;c) =& \phi_A(l_A+c, l_B-c) - \phi_A(l_A, l_B) \label{eq:export}
\end{align}
Equation~\eqref{eq:base} represents an unavoidable shortfall probability that arises even if the capacity $c$ is fully used to supply system A; \eqref{eq:import} is the additional probability of a shortfall due to imports from system B not being available, and \eqref{eq:export} is the probability of a shortfall due to forced exports to system B. The four policies combine the contributions as follows to determine the overall shortfall probability for system A as follows. 
\begin{equation}
p_A(l;c) = \left\{\begin{array}{ll} 
p_{A}^{\textrm{base}}(l;c) + p_{A}^{\textrm{imp}}(l;c), & \text{veto} \\
p_{A}^{\textrm{base}}(l;c) + p_{A}^{\textrm{imp}}(l;c) + p_{A}^{\textrm{exp}}(l;c), & \text{share}\\
p_{A}^{\textrm{base}}(l;c), & \text{assist A} \\
p_{A}^{\textrm{base}}(l;c) + p_{A}^{\textrm{imp}}(l;c) + p_{A}^{\textrm{exp}}(l;c), & \text{assist B} 
\end{array}\right. \label{eq:computerho}
\end{equation}
Definitions \eqref{eq:defrho}-\eqref{eq:computerho} have system A as their reference; results for system B are obtained by exchanging the A and B labels. 

\subsection{Interconnector availability}
The calculations above were performed for an interconnector of fixed capacity $c$. These elementary contributions are easily combined if we assume that interconnection availability is independent of other processes. Let the available interconnection capacities take a set of $n$ discrete values $c_{1:n}$ with probabilities $q_{1:n}$. Then, using additivity of probabilities and \eqref{eq:defrho}, the risk \eqref{eq:rplus} can be computed as
\begin{equation}
r^+_x(l) = h \times \sum_{i=1}^{n} q_i p_x(l;c_i).
\end{equation}

\subsection{Implementation details}

The probability density functions were discretized to probability mass functions for convolutions. A 10~MW grid was used for single-area available generation capacity, and a $50\times 50$~MW grid for all two-dimensional distributions. Linear interpolation was used to distribute off-grid probability mass across neighbouring grid points. The convolution \eqref{eq:convolve} was implemented using a 2D FFT method. For the computation of integrals \eqref{eq:integralphi}, probability masses were assumed to represent a piecewise constant probability density with $50\times50$~MW squares centered on the grid points. All calculations were performed using Python 3.7 running on an Intel i5-7360U CPU under macOS 10.14.1. Run times for all calculations were in the order of seconds, so that speed was not a practical concern.

\section{Model system}

The methodology defined in previous sections is applied to a model system, also used in \cite{Watson2018}, that qualitatively represents the Great Britain system and its interconnection to mainland Europe. The properties of the two systems are based on those of Great Britain (GB) and France (FR). The French system demand is scaled to 150\% of its nominal size to represent the collective interconnection of Great Britain with France as well as other continental European countries. Those countries are effectively treated as a single system, under the assumption of strong interconnections between them. Internal transmission constraints within the GB and FR systems are not taken into account. Historical electricity demand data from GB and France (5 years: 2010-2014) was used to specify an empirical joint probability distribution for demand $D$ in both systems. Net demand measurements were used, which exclude exports and recharging of storage units, and correct for (estimated) output from embedded renewable generation. 

Demand and wind power output were assumed to be statistically independent for this analysis. GB wind power output for the period 2010-2014 was synthesized on the basis of an assumed installed capacity of 13~GW and a capacity factor time series. The capacity factor data (courtesy of Iain Staffell and Kate Ward) was derived from MERRA reanalysis data for wind speeds and an assumed constant distribution of wind generation sites \cite{Staffell2016UsingOutput}. The resulting distribution of GB wind power outputs was used to generate a dependent distribution for wind power in the French system by assuming that both have the same marginal distribution (with 15~GW installed capacity in the scaled French system), and the joint distribution is represented by a Gaussian copula, computed using the $h(\cdot)$-function from \cite[Section C.1]{Aas2009Pair-copulaDependence}. This provides distribution of wind power with an adjustable correlation parameter $\rho$ that varies from 0 (no correlation) to 1 (full correlation). Actual wind power output from the GB and French grid (2014 calendar year) was used to determine a best fit parameter of $\rho=0.5376$. 

In line with typical generation adequacy calculations, dispatchable generators were modelled as independent two-state units that are characterized by their (maximum) capacity and their average availability. Generic unit capacities were used to capture the range of available units: 1200MW for nuclear units, 600MW for large coal/CCGT units, 300MW for smaller units and large hydro units, 150MW for peaking plant, and 80/20/10MW for various smaller units.  A typical unit availability of 90\% was assumed. 

Both Great Britain and France operate capacity markets to maintain their security of supply according to a standard of 3 hours LOLE. We assume that a capacity market guarantees that sufficiently many generators are built to ensure that the standard is met. To maintain a balanced portfolio, units are added in sets that are approximately characteristic of each system (numbers in MW):
\begin{itemize}
    \item GB: set of 1200, 2x600, 2x300, 150, 80, 2x20, 3x10.
    \item FR: set of 2x1200, 600, 300, 150, 80, 2x20, 3x10. 
\end{itemize}
The difference in unit sizes reflects the greater reliance on nuclear units in the French system (resulting in correspondingly larger fluctuations in available capacity). For each system, the number of generator sets was incremented until the LOLE was reduced below 3 hours/year. Then, constant load offsets were determined to achieve a LOLE equal to 3 hours/year (in the absence of interconnection), resulting in:
\begin{itemize}
    \item GB system: 19 generator sets, 649 MW load offset 
    \item FR system: 45 generator sets, 2188 MW load offset
\end{itemize}
The two systems were connected electrically by four interconnectors with independent availabilities of 95\%.

\begin{figure}[!t]
\centering
\includegraphics{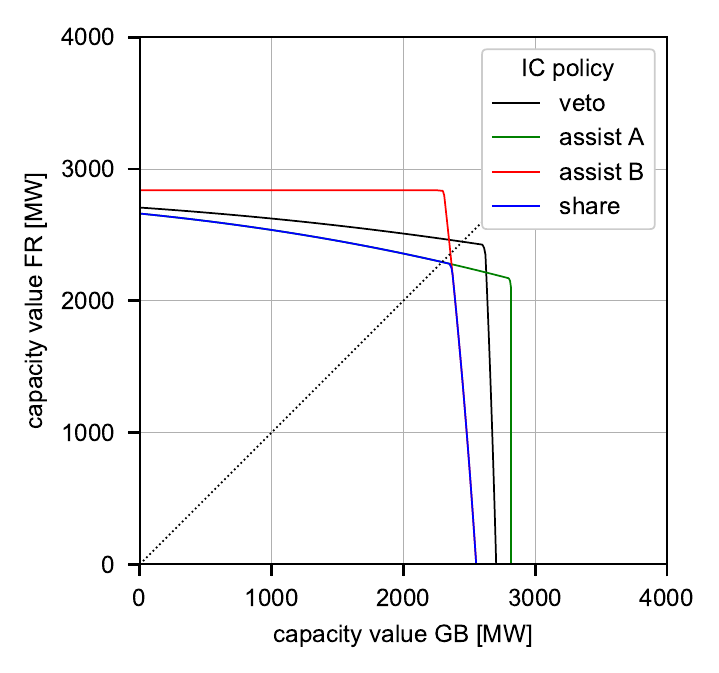}
\caption{Impact of power flow policies on the capacity allocation curve. The \emph{share} curve overlaps with sections of the \emph{assist GB} and \emph{assist FR} curves. The diagonal dotted line is included for reference.}
\label{fig:multipolicy}
\end{figure}

\section{Results}

Figure~\ref{fig:multipolicy} shows capacity allocation curves for the different interconnector policies, for an interconnection capacity that totals 3~GW. The \emph{assist A} and \emph{assist B} policies result in the largest possible capacity contributions to their respective systems, whereas the \emph{veto} policy strikes a balance by restricting exports. The \emph{share} policy results in the smallest capacity values. Because shortfalls are shared between both areas, this policy effectively maximizes the number of loss-of-load instances in each area (albeit at smaller deficit levels). This LOLE-maximising property is also evident from \eqref{eq:computerho}, where all three terms contribute to the risk.  This suggests that the \emph{share} policy can be used as a conservative estimate for the capacity value of interconnection, in the absence of specific power exchange models. 

\begin{figure}[!t]
\centering
\includegraphics{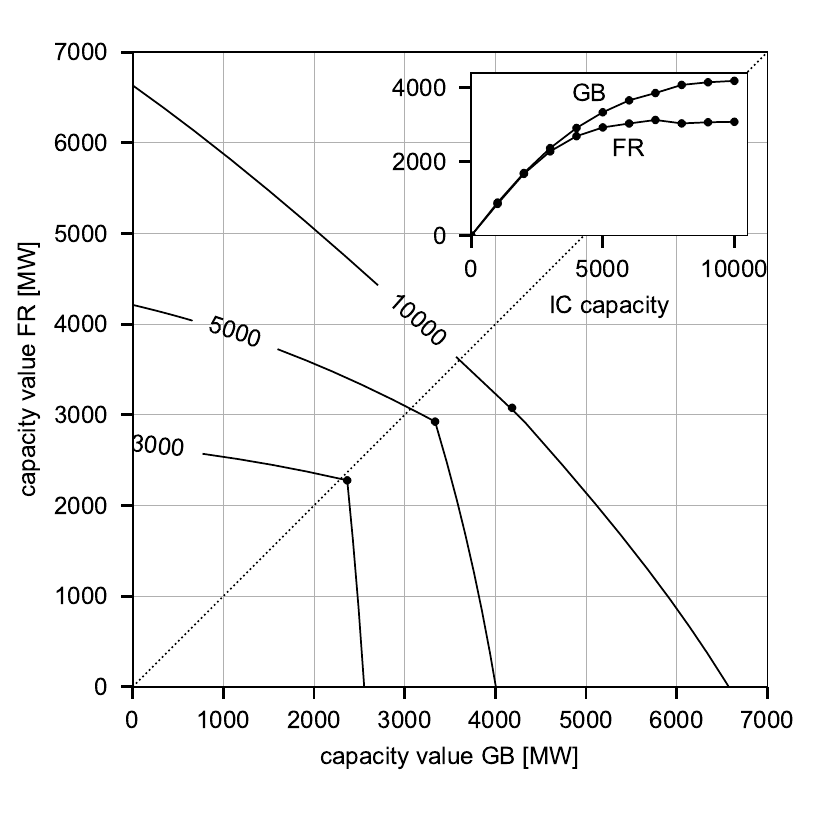}
\caption{\textbf{Main:} Capacity allocation curves for different amounts of aggregate interconnection capacity (3~GW, 5~GW and 10~GW) under the \emph{share} policy. Large dots on the curves indicate the allocations where the joint (GB+FR) capacity value is maximized. \textbf{Inset:} GB and FR capacity values at the maximum contribution point, as a function of installed capacity.}
\label{fig:allocation}
\end{figure}

Figure~\ref{fig:allocation} depicts capacity allocation curves for increasing interconnection capacities of 3~GW, 5~GW and 10~GW. As expected, the capacity value increases with installed capacity, but qualitative changes occur in the shape of the capacity allocation curves. The sharp corner of the 3~GW case implies that an almost-equal capacity benefit befalls both systems with a convincing optimal solution (the corner point). However, the profile of the 10~GW curve is nearly flat, reflecting a direct trade-off between capacity benefits for the GB and FR systems. Whereas insular approaches to capacity assessment are approximately valid for small interconnectors, coordination between capacity mechanisms thus becomes essential when interconnection capacity increases. 

It is worth noting that the allocation curves are asymmetric. Specifically, the point where the total capacity contribution (GB+FR) is maximized –- the corner point in each curve where each area has an LOLE of 3 hours/year –- has a bias towards the smaller GB system: reliance on the interconnector would displace more generation in the GB system than in the FR system, resulting in asymmetric benefits. The inset shows the difference in saturation of the GB and FR capacity values when the capacity value is allocated according to the total-capacity-maximizing approach.

\section{Conclusions and future work}

We have provided a definition for the capacity value of interconnection between neighbouring systems, which takes the form of a capacity allocation curve; an implicit or explicit allocation mechanism is required to select an appropriate value. This curve is sensitive to the operational policy, the most conservative of which is the \emph{share} policy (with respect to the LOLE measure). Even under this conservative policy, the capacity benefits to both neighbouring systems are significant. As interconnection capacities increase, the capacity allocation curves undergo qualitative changes as the capacity benefits saturate, and asymmetric benefits may occur.

In future work, the results presented here can be generalized to other risk measures (e.g. EENS) and capacity value metrics based on Equivalent Firm Capacity (EFC). Capacity values can also be computed for incremental changes in interconnection capacity. For real-world applications, it will be beneficial to extend the proposed framework to handle interconnected systems with more than two areas of interest. This will likely necessitate the use of Monte Carlo algorithms, which are also suitable for time-resolved studies.

%\section*{Acknowledgment}

%This work has received funding from the UK Energy Research Centre (UKERC) as part of the Future Energy System Pathways programme. 



%\bibliographystyle{IEEEtran}
%\bibliography{references,IEEEtranCTL}
%\begin{thebibliography}{1}
%\end{thebibliography}


\end{document}